\documentclass{article}
\usepackage{amsmath}
\newcommand{\bv}{\begin{vmatrix}}
\newcommand{\ev}{\end{vmatrix}}

\newcommand{\bea}{\begin{eqnarray*}}
\newcommand{\eea}{\end{eqnarray*}}
\newcommand{\bean}{\begin{eqnarray}}
\newcommand{\eean}{\end{eqnarray}}

\newcommand{\eqs}[1]{Eqs. (\ref{#1})}
\newcommand{\eq}[1]{Eq.(\ref{#1})}
\newcommand{\meq}[1]{(\ref{#1})}

\newcommand{\ppa}[2]{\left(\frac{\partial}{\partial #1}\right)^{#2}}

\newcommand{\andd}{\ \ \ \textrm{and} \ \ \ }
\newcommand{\eqn}{&=&}
\newcommand{\non}{\nonumber \\}
\newcommand{\sect}[1]{section \ref {#1}}

\title{Five-dimensional Myers-Perry Black Holes as Particle Accelerators  }

\author{Jincheng An\thanks{Email: 201521140011@mail.bnu.edu.cn} and Sijie Gao\thanks{Corresponding author. Email: sijie@bnu.edu.cn} \\
Department of Physics, Beijing Normal University,\\
Beijing 100875, China}

\begin{document}
\maketitle
\begin{abstract}
It has been shown that  black holes could be used as particle accelerators to create arbitrarily high center-of-mass (CM) energy if certain critical conditions are satisfied. Most studies so far are confined in four-dimensional spacetimes. In this paper, we present a systematic analysis on  five-dimensional Myers-Perry black holes and find some novel properties compared to  four-dimensional Kerr black holes. Firstly, we give a rigorous  proof that untrhigh energy collisions cannot occur near a five-dimensional nonextremal black hole. Secondly, For extremal black holes, we find a critical condition on the particles' parameters causing ultraenergetic collisions. Thirdly, when the  spacetime contains a naked singularity, we show that the CM energy could diverge at the singularity if one of the particle just bounces back at the singularity. Finally, we explore a special and important case where the naked singularity just begins to form. Surprisingly, the ultraenergetic collisions do not need any fine-turning in that case. However, we find that one of the conserved angular momentums must be nonzero.

\end{abstract}

\section{Introduction}
Ba\~nados, Silk and West (BSW) proved \cite{BSW} that the center-of-mass(CM) energy of two colliding particles near the event horizon of an extremal Kerr black hole will diverge if one of the particles satisfies certain critical conditions. The BSW mechanism has been further studied and generalized to different spacetimes \cite{vitor}-\cite{mpla}. Some common features have been confirmed for the BSW mechanism. In an extremal black hole, the CM energy could diverge if two particles collide arbitrarily close to the horizon and one of the particles satisfies some critical conditions. No untrahigh-energy collision occurs outside a nonextremal black hole. Ultraenergetic collisions have also been found in spacetimes with naked singularities \cite{nakedk}.

 However, most studies are confined to four-dimensional spacetimes. The purpose of this paper is to investigate ultraenergetic collisions around a five-dimensional spinning black hole, i.e., the Myers-Perry (MP) black hole. This black hole is mainly characterized by two spin parameters $a$ and $b$.  One may expect that the BSW mechanism can be generalized straightforwardly to higher dimensions. However, our work reveals some novel features of the BSW process in higher dimensions. Classified by black hole parameters, our study covers spacetimes containing nonextremal black holes, extremal black holes and naked singularities. As we shall see, the collisions of two particles display different features in different cases.

  The paper is organised as follows. In \sect{sec-mp}, we introduce the general higher dimensional Myers-Perry black holes. In \sect{sec-geod}, the equations of motion of a geodesic in a 5D-MP spacetime are expressed in terms of five conserved quantities.

 In \sect{sec-non}, we prove that collisions in a five-dimensional nonextremal MP spacetime can not produce arbitrarily high CM energy. Although the conclusion is the same as in  four-dimensional Kerr spacetimes, the proof is not trivial. In the 4D case, the constraint from the radial equation is enough to complete the proof. In the 5D case, we find that the angular constraint $\Theta\geq 0$ plays an important role as well. This constraint is usually ignored in the literature.

 In \sect{sec-extre}, we derive the critical condition which causes a divergent CM energy at the extremal horizon. This condition was previously derived for the equal-angular-momenta case $a=b$ \cite{2014}, while we show that it holds for a general extremal black hole.

 In \sect{sec-naked},
we investigate the collision of two ingoing particles in a spacetime containing a naked singularity. We show that if the CM energy diverges, one particle must satisfy a critical condition which makes it bounce back just at the singularity. Similar process was discovered in four-dimensional Kerr spacetimes \cite{nakedk}. The difference is that in the Kerr spacetime, the spin parameter of the spacetime must be arbitrarily close to that of the extremal black hole and the collision takes place arbitrarily close to what would have been the event horizon in the extremal case. In the five-dimensional MP spacetime, we find that the divergent CM energy could be created for any spacetime possessing naked singularities and the collision takes place arbitrarily close to the singularity. Moreover, the BSW process in the four-dimensional Kerr spacetime with a naked singularity does not require a critical particle, unlike the five-dimensional case.

In \sect{sec-special}, we discuss a special case where $b=0$ and $a=1$. This configuration can be viewed as a transition from the extremal case $a<1$ and naked-singularity case $a>1$. Surprisingly, in this spacetime, divergent CM energy can be obtained near the singularity $x=0$ without fine-tuning conditions. The same result was reported in \cite{India} for particle motions on the $\theta=0$ plane. We find that  this kind of collision is general as long as the particles can approach $x=0$ and the angular momentum $\Psi$ of one particle is nonzero.

\section{Myers-Perry Black Holes} \label{sec-mp}
Myers and Perry generalized the four-dimensional asymptotically flat rotating black holes, namely the Kerr black hole, to arbitrary $ D $ dimensions \cite{1986}.
For even dimensions, i.e.,  $D=2n+2$  with $n\geq 2$, the metric can be written as \cite{MP}
\begin{eqnarray}
ds^2&=&-dt^2+\frac{\mu r}{\Pi F}\left(dt+\sum_{i=1}^{n}a_{i}\mu_{i}^2d\phi_{i}\right)^2+\frac{\Pi F}{\Pi-\mu r}dr^2\nonumber\\
&&+\sum_{i=1}^{n}(a_{i}^2+r^2)(d\mu_{i}^2+\mu_{i}^2d\phi_{i}^2)+r^2d\alpha^2,
\end{eqnarray}
where
\begin{equation}
F=1-\sum_{i=1}^{n}\frac{a_{i}^2\mu_{i}^2}{a_{i}^2+r^2}, \ \ \ \Pi=\prod_{i=1}^{n}(a_{i}^2+r^2),
\end{equation}
with the constraint $\sum_{i=1}^{n}\mu_i^2+\alpha^2=1$.

For odd dimensions, $D=2n+1$   with $n\geq2$, the metric takes the form
\begin{eqnarray}\label{mpo}
ds^2&=&-dt^2+\frac{\mu r}{\Pi F}\left(dt+\sum_{i=1}^{n}a_{i}\mu_{i}^2d\phi_{i}\right)^2+\frac{\Pi F}{\Pi-\mu r}dr^2\nonumber\\
&&+\sum_{i=1}^{n}(a_{i}^2+r^2)(d\mu_{i}^2+\mu_{i}^2d\phi_{i}^2),
\end{eqnarray}
with the constraint $\sum_{i=1}^{n}\mu_i^2=1$.

The constant  $\mu$  is related to the mass of the black hole by
\begin{equation}
M=\frac{(D-2)\Omega_{D-2}}{16\pi G}\mu ,
\end{equation}
where
\begin{equation}
\Omega_{D-2}=\frac{2\pi^{\frac{D-1}{2}}}{\Gamma(\frac{D-1}{2})}.
\end{equation}
is the area of the unit $(D-2)$-sphere.

In addition to some familiar symmetries that can be described by Killing vectors, there are also hidden symmetries in higher dimensional spacetimes,  which are described by Killing  tensors. In Myers-Perry black holes, these tensors can be explicitly written as \cite{MP}
\begin{eqnarray}
K_{b}^{a(k)}&=&\frac{(2k)!}{(2^k k!)^2}\left(\delta_{b}^ah^{[a_{1}b_{1}}...h^{a_{k}b_{k}]}h_{[a_{1}b_{1}}...h_{a_{k}b_{k}]} \right.\nonumber\\
&&\left.-2kh^{a[b_{1}}...h^{a_{k}b_{k}]}h_{b[b_{1}}...h_{a_{k}b_{k}]}\right),
\end{eqnarray}
where
\begin{equation}
h=\sum_{i=1}^{n}a_{i}\mu_{i}d\mu_{i}\wedge\left(a_{i}dt+(a_{i}^2+r^2)d\phi_{i}\right)+r dr\wedge\left(dt+\sum_{i=1}^{n}a_{i}\mu_{i}^2d\phi_{i}\right),
\end{equation}
is called Killing-Yano(CCKY) tensor satisfying
\begin{equation}
\bigtriangledown_{(a}h_{b)c}=\frac{1}{D-1}\left(g_{ab}\bigtriangledown_{d}h_{c}^{d}-
\bigtriangledown_{d}h_{(a}^{d}g_{b)c}\right)\,.
\end{equation}
 $K_{ab}$ satisfies the identity
\begin{equation}
 \bigtriangledown_{(c}K_{ab)}=0.
 \end{equation}
If $u^a$ is the unit tangent of a geodesic, then along the geodesic, one can show that
\bean
c_{k}=K_{ab}^{(k)}u^au^b
\eean
is a constant. For example,  $K_{ab}^{(0)}$ is just the metric  $g_{ab}$, which gives the normalization condition $g_{ab}u^au^b=-1$.
For $D =2n+1$, there are $(n+1)$ Killing vectors and $n$ Killing tensors, while  for $D =2n+2$, there are  $(n+1)$ Killing vectors and $n+1$ Killing tensors.
\section{Timelike Geodesics in a 5-dimensional Myers-Perry black hole}\label{sec-geod}
In this section, we will investigate the geodesic motion of a particle  in a 5-dimensional Myers-Perry spacetime, which is essential for studying collisions of two particles in the following sections. For $ D=5 $ , the metric \eqref{mpo} becomes
\begin{eqnarray}\label{mpr}
ds^2&=&-dt^2+\frac{\mu}{\Sigma}\left(dt+a\sin{\theta}^2d\phi+b\cos{\theta}^2d\psi\right)^2+\frac{r^2\Sigma}{\Pi-\mu r^2}dr^2\nonumber\\
&&+\Sigma d\theta^2+(r^2+a^2)\sin{\theta}^2d\phi^2+(r^2+b^2)\cos{\theta}^2d\psi^2 ,
\end{eqnarray}
where
\begin{eqnarray}
\Sigma&=&r^2+a^2\cos{\theta}^2+b^2\sin{\theta}^2,\\
\Pi&=&(r^2+a^2)(r^2+b^2).
\end{eqnarray}
For convenience, we let $ x=r^2 $ and express \eq{mpr} in the Boyer-Lindquist coordinates\cite{5D} as
\begin{eqnarray}
ds^2&=&\frac{\rho^2}{4\Delta}dx^2+\rho^2d\theta^2-dt^2+(x+a^2)\sin{\theta}^2d\phi^2+(x+b^2)\cos{\theta}^2d\psi^2\nonumber\\
&&+\frac{r_{0}^2}{\rho^2}\left(dt+a\sin{\theta}^2d\phi+b\cos{\theta}^2d\psi\right)^2 ,
\end{eqnarray}
with
\begin{eqnarray}
\rho^2&=&x+a^2\cos{\theta}^2+b^2\sin{\theta}^2,\\
\Delta&=&(x+a^2)(x+b^2)-r_{0}^2x,
\end{eqnarray}
where $r_{0}$ is the length parameter \cite{5DF} related to the mass of the black hole by $M=\frac{3r_{0}^2}{8\sqrt{\pi}G}$, and $G$ is the 5-dimensional gravitational coupling constant. Without loss of generality, we shall take $r_0=1$ in the rest of the paper.

This  spacetime  possesses three Killing vectors  $ \xi^{a} =(\frac{\partial}{\partial{t}})^{a} $ ,$\phi^{a} =(\frac{\partial}{\partial{\phi}})^{a} $ and $ \psi^{a} =(\frac{\partial}{\partial{\psi}})^{a} $. There are also two Killing  tensors, one of which is just the metric $ g_{ab} $. The other one is given by \cite{MP}
\begin{equation}
K_{ab}=-h_{a}^{c}h_{bc}+\frac{1}{2}g_{ab}h_{cd}h^{cd},
\end{equation}
which, in the Boyer-Lindquist coordinates, takes the form
\begin{equation}
 K^{\mu\nu}=-(a^2\cos{\theta}^2+b^2\sin{\theta}^2)(g^{\mu\nu}+\delta_{t}^\mu\delta_{t}^\nu)
 +\frac{\delta_{\phi}^\mu\delta_{\phi}^\nu}{\sin{\theta}^2}+\frac{\delta_{\psi}^\mu
 \delta_{\psi}^\nu}{\cos{\theta}^2}+\delta_{\theta}^\mu\delta_{\theta}^\nu.
\end{equation}
Let
\begin{eqnarray}
v^{a}=\ppa{\tau}{a}=\dot x^\mu \ppa{x^\mu}{a}\,,
\end{eqnarray}
be the ``5-velocity'' of a particle moving along a geodesic, where $\tau$ is the proper time. Then the Killing vectors and tensors give rise to five conserved quantities
\begin{equation}\label{n1}
g_{ab}v^{a}v^{b}=-1 ,
\end{equation}
\begin{equation}\label{Ec}
-g_{ab}\xi^{a}v^{b}=E ,
\end{equation}
\begin{equation}\label{Lf}
g_{ab}\phi^{a}v^{b}=\Phi,
\end{equation}
\begin{equation}\label{Lp}
g_{ab}\psi^{a}v^{b}=\Psi,
\end{equation}
\begin{equation}\label{Kc}
K_{ab}v^{a}v^{b}=K.
\end{equation}
By solving these  equations, one can obtain the following first-order differential equations \cite{5D}
\begin{eqnarray}\label{eqm}
\dot{t}&=&\frac{(x+a^2)(x+b^2)}{\rho^2\Delta}\epsilon+E,\\
\dot{x}&=&\pm\frac{2\sqrt{\Xi}}{\rho^2}, \label{xdot}\\
\dot{\theta}&=&\pm\frac{\sqrt{\Theta}}{\rho^2}, \label{thetadot}\\
\dot{\phi}&=&\frac{\Phi}{\rho^2\sin{\theta}^2}-\frac{a(x+a^2)}{\rho^2\Delta}\epsilon-\frac{(a^2-b^2)\Phi}{(x+a^2)\rho^2},\\\label{phd}
\dot{\psi}&=&\frac{\Psi}{\rho^2\cos{\theta}^2}-\frac{b(x+b^2)}{\rho^2\Delta}\epsilon+\frac{(a^2-b^2)\Psi}{(x+b^2)\rho^2},\label{psd}
\end{eqnarray}
where
\begin{eqnarray}\label{eqm1}
\epsilon=E+\frac{a\Phi}{x+a^2}+\frac{b\Psi}{x+b^2} ,
\end{eqnarray}
\begin{eqnarray}\label{Theta}
\Theta=(E^2-m^2)(a^2\cos{\theta}^2+b^2\sin{\theta}^2)-\frac{\Phi^2}{\sin{\theta}^2}-\frac{\Psi^2}{\cos{\theta}^2}+K ,
\end{eqnarray}
\begin{eqnarray}\label{xi}
\Xi&=&\Delta\left[x(E^2-m^2)+(a^2-b^2)\left(\frac{\Phi^2}{x+a^2}-\frac{\Psi^2}{x+b^2}\right)-K\right]\nonumber\\
&&+(x+a^2)(x+b^2)\epsilon^2.
\end{eqnarray}
In the following sections, we will use these results to  calculate the CM energy of two colliding particles.

\section{Collisions near the Horizon of a Non-Extremal 5-dimensional Myers-Perry Black Hole}\label{sec-non}
The horizons of the MP black hole are located at $\Delta=0$. The solutions are given by
\begin{eqnarray}\label{ghx}
x=x_{\pm}=\frac{1}{2}\left[1-a^2-b^2\pm\sqrt{(1-a^2-b^2)^2-4a^2b^2}\right]\,.
\end{eqnarray}
where $x=x_+$ is the position of the event horizon.
A nonextremal black possesses two horizons, which requires
\bean
|1-a^2-b^2|>2|ab|\,.
\eean

Consider the collision of two particles with the same mass $m$ near the event horizon. The energy of center of mass \cite{BSW} is given by
\begin{equation}
E_{cm}=\sqrt{2}m\sqrt{1-g_{ab}v_{1}^av_{2}^b}.
\end{equation}
Because our purpose is to check whether $E_{cm}$ could be divergent, we define the effective CM energy \cite{4D} as
\begin{equation}\label{Ef}
E_{eff}=-g_{ab}v_{1}^av_{2}^b.
\end{equation}
 To simplify the calculation, we also assume that the collision takes place at $\theta = \frac{\pi}{4}$. However, we shall see that our results are independent of this choice. Then $E_{eff}$ can be written in the form
\begin{eqnarray}\label{eff}
E_{eff}=-2\rho^2\dot\theta_1\dot\theta_2+ \frac{Nu}{2\Delta\rho^2},
\end{eqnarray}
Obviously, the first term on the right-hand side of \eq{eff} can not be divergent and the second term suggests that a possible divergent CM energy can only be obtained at the horizon $\Delta=0$. To see if the divergence can actually happen, we  expand the $Nu$ at $x=x_+$ as
\bean
Nu=\alpha_0+\alpha_1(x-x_+)+\alpha_2(x-x_+)^2+...\,,
\eean
and then $E_{eff}$ is in the form
\bean
E_{eff}\sim \frac{\alpha_0+\alpha_1(x-x_+)+\alpha_2(x-x_+)^2+...}{x-x_+}\,. \label{eesim}
\eean

 Since the geodesic is future-directed, one can show that outside the horizon
\begin{eqnarray}
\dot{t}=\frac{[(x+a^2)(x+b^2)-1]\epsilon+E\rho^2\Delta}{\rho^2\Delta}\geq 0 .
\end{eqnarray}
 By expanding the numerator at the horizon $x=x_+$ and requiring the leading term to be non-negative, we get
\begin{eqnarray}\label{resg}
(1-a^2-b^2+J)E+a(1-a^2+b^2+J)\Phi+b(1+a^2-b^2+J)\Psi\geq 0\,,
\end{eqnarray}
where
\bean
J=\sqrt{(1-a^2-b^2)^2-4a^2b^2}\,.
\eean
Next, we calculate $Nu$ in \eqref{eff} at the horizon, which yields
\begin{eqnarray}\label{ga0}
Nu\mid_{x=x_+}&=&(1-a^2-b^2+J)E_1E_2\nonumber\\
&&+\left[a^2-a^4+b^2+2a^2b^2-b^4+(a^2-b^2)J\right]\Phi_1\Phi_2\nonumber\\
&&+\left[a^2-a^4+b^2+2a^2b^2-b^4-(a^2-b^2)J\right]\Psi_1\Psi_2\nonumber\\
&&+a(1-a^2+b^2+J)(E_1\Phi_2+E_2\Phi_1)\nonumber\\
&&+b(1+a^2-b^2+J)(E_1\Psi_2+E_2\Psi_1)\nonumber\\
&&+2ab(\Phi_1\Psi_2+\Phi_2\Psi_1)-\sqrt{I_1}\sqrt{I_2},
\end{eqnarray}
where
\begin{eqnarray}
I=\frac{\left[(1-a^2-b^2+J)E+a(1-a^2+b^2+J)\Phi+b(1+a^2-b^2+J)\Psi\right]^2}{1-a^2-b^2+J}.
\end{eqnarray}
With the help of \eq{resg}, one can show that $ Nu\mid_{x=x_+} = 0 $, which means
\bean
\alpha_0=0\,.
\eean

Next, we need to calculate $\alpha_1$ because by l'Hospital's rule, $E_{eff}$ in \eq{eesim} is divergent at the horizon if $\alpha_1$ is divergent.
Since the explicit expression of $\alpha_1$ for a general non-extremal black hole is very lengthy and complicated, we will concentrate on two special cases , the $ Singly $ $ Rotating  $ case when $b = 0$  and the $ Equally $ $ Doubly $ $ Rotating $  case when $a = b $. In both cases, we shall show that $\alpha_1$ can be infinite supposing some conditions are imposed to the particles. But these conditions just prevent the particles from reaching the horizon.
\subsection{Singly Rotating Black Holes  with $ b=0 $ }
In this subsection, we explore the singly rotating black holes where $a \neq 0$ and $ b =0 $. The event horizon is located at
\begin{eqnarray}
x=x_+= 1-a^2 .
\end{eqnarray}
Without loss of generality, we can set $ a > 0 $. The non-extremal  condition then becomes  $0 < a < 1$ . The restriction \eqref{resg} reduces to
\begin{equation}
E+a\Phi\geq 0 . \label{eal}
\end{equation}
 $Nu$ in \eq{eff} becomes
\begin{eqnarray}
Nu&=&a^2xE_1E_2+a^4xE_1E_2+3a^2x^2E_1E_2+2x^3E_1E_2+2axE_2\Phi_1\nonumber\\
&&+2axE_1\Phi_2+4x\Phi_1\Phi_2-2a^2x\Phi_1\Phi_2-4x^2\Phi_1\Phi_2\nonumber\\
&&+(2a^2-2a^4+4x-6a^2x-4x^2)\Psi_1\Psi_2\nonumber\\
&&-2\sqrt{G_1}\sqrt{G_2}+x[(1-a^2)-x]\sqrt{H_1}\sqrt{H_2},
\end{eqnarray}
where
\begin{eqnarray}
G&=&x\left[x-(1-a^2)\right]\left[x(E^2-m^2)-K+a^2\left(\frac{\Phi^2}{a^2+x}-\frac{\Psi^2}{x}\right)\right]\nonumber\\
&&+x(a^2+x)\left(E+\frac{a\Phi}{a^2+x}\right)^2 ,
\end{eqnarray}
and
\begin{eqnarray}
H=a^2E^2-a^2m^2+2K-4\Phi^2-4\Psi^2  .
\end{eqnarray}

We have shown that $\alpha_0$ in \eq{eesim} vanishes. Straightforward calculation yields
\begin{eqnarray}\label{a1s}
\alpha_1&=&(6-5a^2+a^4)E_1E_2+(6a^2-4)\Phi_1\Phi_2\nonumber\\
&&+2(a^2-2)\Psi_1\Psi_2+2a(E_2\Phi_1+E_1\Phi_2)\nonumber\\
&&-(1-a^2)\sqrt{H_1}\sqrt{H_2}+\frac{E_2+a\Phi_2}{E_1+a\Phi_1}T_1+\frac{E_1+a\Phi_1}{E_2+a\Phi_2}T_2,
\end{eqnarray}
where
\begin{eqnarray}
T=(1-a^2)m^2-(3-3a^2+a^4)E^2+K-2aE\Phi-a^2(K+\Phi^2-\Psi^2).
\end{eqnarray}
Now it is obvious that an infinite $\alpha_1$ is possible only if one of the particles satisfies
\bean\label{EaL0}
E+a\Phi=0.
\eean
Although this relation was derived by setting $\theta=\frac{\pi}{4}$, one can check that it also holds for other values of $\theta$. In the rest of this subsection, we will discuss collisions at a general value of $\theta$.

We have noticed that the collision must occur at the horizon (accurately speaking, it occurs arbitrarily close to the horizon). So we need to check whether the particle can reach the horizon under the critical relation \meq{EaL0}. It follows from \eqs{xdot} and \meq{thetadot} that
\bean
\Xi|_{x=x_+}\geq 0 \,,
\eean
and
\bean
\Theta|_{x=x_+}\geq 0  \,,\label{btb}
\eean
must hold.
By substituting \eq{EaL0} into \eq{xi}, we find
\begin{eqnarray}\label{co}
\Xi\mid_{x=x_+}=0 \text{ },
\end{eqnarray}
This result is the same as in the four-dimensional case. Since $\Xi$ must be nonnegative in a neighborhood outside the horizon \cite{4D}, we have
\begin{eqnarray}\label{dx}
\frac{\partial\Xi}{\partial x}\Big|_{x_+}=(2-3a^2+a^4)E^2-(a^2-1)^2m^2+(a^2-1)K-a^2\Psi^2\geq0\,.
\end{eqnarray}
This inequality alone can be easily satisfied by choosing appropriate parameters. However, we show now that it is inconsistent with \eq{btb}.
By substituting \eq{EaL0} and $b=0$  into \eq{Theta}, we find
\bean
\Theta=K+a^2(E^2-m^2)\cos{\theta}^2-\frac{E^2}{a^2\sin{\theta}^2}-\frac{\Psi^2}{\cos{\theta}^2}.
\eean
Together with inequality \meq{dx}, we have
\bean\label{iqb=0}
\Theta&\leq&\frac{1}{a^2(1-a^2)\cos{\theta}^2\sin{\theta}^2}\big[(a^2-1)(1-a^2\sin{\theta}^2)^2
\cos{\theta}^2E^2-a^2 \sin{\theta}^2\nonumber\\
&&(1-a^2)\sin{\theta}^2\Psi^2 -a^2(1-a^2)(1-a^2\sin{\theta}^2)\cos{\theta}^2\sin{\theta}^2m^2\big]\,.
\eean
Since $0<a^2<1$ for a non-extremal black hole, it's easy to check that the coefficients of $E^2$, $\Psi^2$ and $m^2$ in the numerator are all negative. Thus, \eq{btb} fails to hold.

This means that an infinite CM energy can not be obtained because the particle with the critical energy cannot reach the horizon of the nonextremal black hole.

\subsection{Doubly Equally Rotating Black Hole where $a = b $}
In the previous subsection, we set the spinning parameter $b=0$. Note that the five-dimensional MP black hole possesses two spinning parameters, $a$ and $b$, differing from the four-dimensional cases. So it would be interesting to check the cases where both $a$ and $b$ are nonzero. In this subsection, we set $a = b\neq 0 $.  Now the event horizon $x_{+}$ reduces to
\begin{equation}
x_{+}=\frac{1-2a^2+\sqrt{1-4a^2}}{2},
\end{equation}
where $0<a<\frac{1}{2}$, as required by the  non-extremal condition. The relation \eqref{resg} becomes
\begin{eqnarray}
(1+\sqrt{1-4a^2})E+2a(\Phi+\Psi)\geq0.
\end{eqnarray}
After taking $ b=a $, $ Nu$  in \eqref{eff} becomes
\begin{eqnarray}
Nu&=&2(a^4+a^6+a^2x+3a^4x+3a^2x^2+x^3)E_1E_2\nonumber\\
&&+2a^3E_2\Phi_1+2axE_2\Phi_1+2a^3E_1\Phi_2+2axE_1\Phi_2\nonumber\\
&&+2(a^2-2a^4+2x-4a^2x-2x^2)\Phi_1\Phi_2\nonumber\\
&&+2a^3E_2\Psi_1+2axE_2\Psi_1+2a^3E_1\Psi_2+2axE_1\Psi_2\nonumber\\
&&+2(a^2-2a^4+2x-4a^2x-2x^2)\Psi_1\Psi_2\nonumber\\
&&+2a^2(\Psi_1\Phi_2+\Phi_1\Psi_2)-2\sqrt{Q_1}\sqrt{Q_2}\nonumber\\
&&-2(a^4+2a^2x+x^2-x)\sqrt{O_1}\sqrt{O_2},
\end{eqnarray}
where
\begin{equation}
O=-a^2m^2+a^2E^2+K-2\Phi^2-2\Psi^2,
\end{equation}
\begin{eqnarray}
Q&=&[(a^2+x)^2-x](-m^2x+xE^2-K) \non
&+&(a^2+x)[(a^2+x)E+a(\Phi+\Psi)]^2.
\end{eqnarray}
After expanding  $ Nu $ at horizon, we find
\begin{eqnarray}\label{a1d}
\alpha_1&=&[4a^2-3(1+\sqrt{1-4a^2})]E_1E_2+4\sqrt{1-4a^2}(\Phi_1\Phi_2+\Psi_1\Psi_2)\nonumber\\
&&-2a[E_1(\Phi_2+\Psi_2)+E_2(\Phi_1+\Psi_1)]\nonumber\\
&&+\frac{P_2}{P_1}\left\{E_1P_1-\frac{1}{4}\sqrt{1-4a^2}\left[\left(1+\sqrt{1-4a^2}\right)^2\left(m^2-E_1^2\right)-4K_1\right]\right\}\nonumber\\
&&+\frac{P_1}{P_2}\left\{E_2P_2-\frac{1}{4}\sqrt{1-4a^2}\left[\left(1+\sqrt{1-4a^2}\right)^2\left(m^2-E_2^2\right)-4K_2\right]\right\}\nonumber\\
&&-2\sqrt{1-4a^2}\sqrt{P_1}\sqrt{P_2},
\end{eqnarray}
where
\begin{eqnarray}
P=(1+\sqrt{1-4a^2})E+2a(\Phi+\Psi).
\end{eqnarray}
Then it is clear that $P=0$ must hold for one of the particles to make $\alpha_1 $ divergent. Using the approach similar to that in the previous subsection, one can show that $\Xi\geq0$ and $\Theta\geq0$ cannot both hold outside the horizon if  $P=0$. This, again, means that the CM energy cannot be divergent for nonextremal black holes.

\section{Collisions near the Horizon of an Extremal Myers-Perry Black Hole } \label{sec-extre}
Now consider the extremal Myers-Perry black holes, i.e., $ x_-=x_+ $ in \eq{ghx}. To be specific, we choose
\begin{eqnarray}
b=1-a\,.
\end{eqnarray}
Then the extremal horizon is located at
\begin{eqnarray}\label{eh}
x=x_+=a(1-a)\,.
\end{eqnarray}
with $0\leq a\leq 1$. However, the calculation for $a=0$ or $a=1$ is quite different, as we shall see in \sect{sec-naked}. Thus, we shall restrict ourselves to
\bean
0<a<1
\eean
in this section.
Now the relation \eqref{resg} reduces to
\begin{equation}\label{rese}
\Phi+\Psi+E\geq0.
\end{equation}

Because  $\Delta=(x-x_+)^2$, \eq{eesim} is modified as
\bean
E_{eff}\sim \frac{\alpha_0+\alpha_1(x-x_+)+\alpha_2(x-x_+)^2+...}{(x-x_+)^2} \label{eex}
\eean

Similarly, denote the numerator of \eq{eex} by $Nu$ and set $\theta=\frac{\pi}{4}$ at the collision point. We find that
\begin{eqnarray}
Nu&=&\big(3a^2-8a^3+9a^4-6a^5+2a^6+2x+12a^2x-12a^3x\nonumber\\
&&+6a^4x+3x^2-6ax^2+6a^2x^2+2x^3\big)E_1E_2\nonumber\\
&&+\big(4a-12a^2+12a^3-4a^4-2x-12ax-8a^2x-4x^2\big)\Phi_1\Phi_2\nonumber\\
&&+\big[4a^3-4a^4+4ax-8a^2x+2(1-2x)x\big]\Psi_1\Psi_2\nonumber\\
&&+2a\big[(1-a)^2+x\big](E_1\Phi_1+E_2\Phi_2)\nonumber\\
&&+2(1-a)(a^2+x)(E_1\Psi_1+E_2\Psi_2)-\Delta\sqrt{X_1}\sqrt{X_2}\nonumber\\
&&+2(1-a)a(\Phi_1\Psi_2+\Phi_2\Psi_1)-2\sqrt{Y_1}\sqrt{Y_2},
\end{eqnarray}
where
\begin{eqnarray}
X=(1-2a+2a^2)(E^2-m^2)+2K-4\Phi^2-4\Psi^2,
\end{eqnarray}
\begin{eqnarray}
Y&=&(a^2+x)(1-2a+a^2+x)\left[E+\frac{a\Phi}{a^2+x}+\frac{(1-a)\Psi}{1-2a+a^2+x}\right]^2\nonumber\\
&&+\Delta\left[x(E^2-m^2)-K-(1-2a)\left(\frac{a\Phi}{a^2+x}+\frac{(1-a)\Psi}{1-2a+a^2+x}\right)\right].
\end{eqnarray}
 After expanding  $ Nu$ as before, using the restriction \eqref{rese}, we find that
\bean
\alpha_0=\alpha_1=0 \,.
\eean
Due to the quadratic denominator in \eq{eex}, we need to calculate $\alpha_2$, which gives
\begin{eqnarray}\label{a2}
\alpha_2&=&3E_1E_2-4(\Phi_1\Phi_2+\Psi_1\Psi_2)-\sqrt{X_1}\sqrt{X_2}\nonumber\\
&&-\frac{[E_1+(2a-1)(\Phi_1-\Psi_1)][E_2+(2a-1)(\Phi_2-\Psi_2)]}{2a(1-a)}\nonumber\\
&&+\frac{(E_1+\Phi_1+\Psi_1)U_2}{4a(1-a)(E_2+\Phi_2+\Psi_2)}+\frac{(E_2+\Phi_2+\Psi_2)U_1}{4a(1-a)(E_1+\Phi_1+\Psi_1)},
\end{eqnarray}
where
\begin{eqnarray}
U&=&4(a-1)a\big[a(a-1)m^2+(1+a-a^2)E^2-K\big]\nonumber\\
&&+[E+(2a-1)(\Phi-\Psi)]^2\,.
\end{eqnarray}
It is then clear that as long as one of the two particles satisfies the critical relation
\begin{eqnarray}\label{cre}
\Phi+\Psi+E =0\,,
\end{eqnarray}
$\alpha_2$ will blow up, causing the divergence of the CM energy.
Similar to the nonextremal case, we find that \eq{cre} is independent of the value of $\theta$. So \eq{cre} is a general critical relation for extremal black holes.

We also need to check whether the particle satisfying the critical condition can reach the horizon. \eqs{xdot} and \meq{thetadot} imply that both $\Xi$ and $\Theta$ should be nonnegative in a neighborhood of the horizon. By substitution of \eq{cre}, we obtain
\begin{eqnarray}\label{exx}
\Xi\eqn (E_c^2-m^2)[x-(1-a)a]^2\left(x-\frac{K-E_c^2}{E_c^2-m^2}\right)\,, \\
\Theta\eqn K-\frac{\Phi^2}{\sin{\theta}^2}-\frac{\Psi^2}{\cos{\theta}^2}+
(E_c^2-m^2)\left[a^2\cos{\theta}^2+(1-a)^2\sin{\theta}^2\right]\,.
\end{eqnarray}
where $E_c=-\Phi-\Psi$. Obviously, if we choose
\bean
E_c^2>m^2, \ \ \textrm{and} \ \ \ K<E_c^2\,,
\eean
$\Xi$ will be positive everywhere. If one choose the parameters more carefully, the positivity of $\Theta$ near the horizon can also be guaranteed. Hence, unlike the nonextremal case we have discussed, the divergent CM energy can be realized for extremal black holes.

\section{Collisions in a spacetime admitting  naked singularity } \label{sec-naked}
In this section, we will study the collision in a 5-dimensional Myers-Perry spacetime with naked singularity.
For convenience, we set $\mu=1$ in \eq{mpr} and then  obtain the $Kretchman $ invariant as \cite{MP}
\begin{eqnarray}
R_{abcd}R^{abcd}=\frac{24}{\Sigma^6}\big(4x-3\Sigma\big)\big(4x-3\Sigma\big).
\end{eqnarray}
In particular, at $ x=0 $, we have
\begin{eqnarray}
R_{abcd}R^{abcd}\big|_{x=0}=\frac{72}{(a^2\cos{\theta}^2+b^2\sin{\theta}^2)^4}\,.
\end{eqnarray}
Hence a singularity occurs at
\bean
x=0,\ \ \textrm{and} \ \ \theta=\frac{\pi}{2}
\eean
if
\bean
b=0
\eean
In this case, the event horizon is located at
\begin{eqnarray}
x_H=1-a^2,
\end{eqnarray}

It is obvious that the singularity is naked when $a>1$. We shall show that infinite CM energy could be produced when a naked singularity is present.

Now we calculate collisions for the spacetime with $b=0$ and $a>1$. If the orbit is in the $\theta=\pi/2$ plane, \eq{Theta} indicates that $\Psi$ must vanish and
\bean
K=\Phi^2\,.
\eean
Then $E_{eff}$ takes the simple form
\begin{eqnarray}\label{effn}
E_{eff}&=&\frac{1}{x(x+a^2-1)}\left[(aE_1+\Phi_1)(aE_2+\Phi_2)+(a^2E_1E_2-\Phi_1\Phi_2)x\right.\nonumber\\
&&+\left.E_1E_2x^2\pm\sqrt{(aE_1+\Phi_1)^2+C_1}\sqrt{(aE_2+\Phi_2)^2+C_2}\right],
\end{eqnarray}
where
\begin{eqnarray}
C=(E^2-m^2)x^2+\big[a^2(E^2-m^2)-\Phi^2+m^2\big]x\,.
\end{eqnarray}
The $-$ sign corresponds to the collision between an ingoing particle ($\dot  x<0$ )and an outgoing particle($\dot x>0$), while the $+$ sign corresponds to the collision between two ingoing particles. This difference will be important in the following analysis.

To make $E_{eff}$ blow up, the collision must occur at $x=0$, as shown in \eq{effn}. It is not difficult to find that the numerator of \eq{effn} vanishes at $x=0$ for two ingoing particles. This is not the case that we should pay attention to. However, for the ``$-$'' sign, i.e., when two particles come from different radial directions, we find
\bean
\lim_{x\rightarrow 0} E_{eff}=\frac{2(aE_1+\Phi_1)(aE_2+\Phi_2)}{x(1-a^2)}\rightarrow \infty  \label{einff}
\eean
This divergence does not seem to involve any critical condition as usual. But we should emphasize that \eq{einff} requires that one particle becomes outgoing at$x=0$! This is equivalent to saying that one particle coming from infinity has a turning point arbitrarily close to $x=0$.

To see if this could happen, we write down the radial equation derived from \eq{xdot} as
\bean
\dot x^2=-V_{eff}(x)\,,
\eean
where
\bean
V_{eff}(x)=-\frac{4\big(E^2-m^2\big)x^2+4\big(E^2a^2-m^2a^2+m^2-\Phi^2\big)x+4\big(aE+\Phi^2\big)  }{x}\,.
\eean
The turning point at $x=0$ simply means $V_{eff}(0)=0$, which yields
\bean
aE+\Phi=0\,.
\eean
This is just the critical condition we are looking for. But it was derived in a subtle manner. The physical picture is that we send a particle with the critical condition from infinity and it just becomes outgoing  at the naked singularity. If it collides with any ingoing particle arbitrarily close to the singularity, the CM energy will diverge.

\section{The special case of $b=0$ and $ a=1$} \label{sec-special}
For $b=0$ and $ a=1$, there is a naked conical singularity at \cite{2014}
\bean
x=0 \andd \theta\neq \frac{\pi}{2}\,,
\eean
and as we have mentioned, there is a naked curvature singularity at $x=0$ and $\theta=\frac{\pi}{2}$.
This can be viewed as a critical configuration connecting the extremal case ($a<1$) in section \ref{sec-extre} and the case of naked singularity ($ a>1$) in section \ref{sec-naked}. We have seen that the analysis in \sect{sec-extre} does not apply to the case $a=1$.  Note that, for $b=0$,
\bean
\lim_{x\rightarrow 0}\Xi\rightarrow a^2(1-a^2)\Psi^2 \,.
\eean
Consequently, if $a>1$, there is no particle which could reach $x=0$ unless $\Psi=0$. This is why we restricted ourselves to $\theta=\pi/2$ and $\Psi=0$ in \sect{sec-naked}. However, for $a=1$, we are free to consider other configurations, particularly the case $\Psi\neq 0$. We consider two ingoing particles colliding somewhere. First, we notice that \eq{eal} becomes
\bean
E+\Phi\geq 0  \label{eepp}
\eean
 To be definitive, we specify $\theta=\pi/3$ at the collision point. Then we find
\bean
E_{eff}=\frac{Nu}{3x(1+4x)}\,,
\eean
where $Nu$ is regular everywhere. Thus, collisions with infinite CM energy can only occur at $x=0$. It is straightforward to find
\bean
\lim_{x\rightarrow 0} Nu\rightarrow 12\left[-(E_1+\Phi_1)(E_2+\Phi_2)+\Psi_1\Psi_2+\sqrt{(E_1+\Phi_1)^2-\Psi_1^2} \sqrt{(E_2+\Phi_2)^2-\Psi_2^2} \right]\,.
\eean
It is obvious, with the help of \eq{eepp}, that $Nu$ vanishes if $\Psi_1=\Psi_2=0$. This is not the case that we are interested in. But if $\Psi_1$ or $\Psi_2$ does not vanish, $E_{eff}$ will be divergent in general. Note that this is a generic divergence without a fine-tuning for either particle. But we still need to be cautious because we do not know yet whether the particles can approach $x=0$, which requires that both $\Xi$ and $\Theta$ are nonnegative.

 For $a=1, b=0$, we find that
\begin{eqnarray}
\Xi\eqn x\left[(E^2-m^2)x^2+(E^2-K)x+(E+\Phi)^2-\Psi^2\right]\,,\\
\Theta\eqn K+(E^2-m^2)\cos{\theta}^2-\frac{\Phi^2}{\sin{\theta}^2}-\frac{\Psi^2}{\cos{\theta}^2}\,,
\end{eqnarray}
Obviously, if we choose
\bean
(E+\Phi)^2>\Psi^2\,,
\eean
and appropriate $K$ and $m$, the positivity of $\Xi$ and $\Theta$ can be guaranteed in a neighborhood of $x=0$.

To see if an orbit can exist globally, we restrict the particle in the $\theta=0$ plane, which leads to $\Phi=0$ \cite{5DF}. Solving $\Theta(\theta=0)=0$ for $K$, we have \cite{5D}
\bean
K=\Psi^2-(E^2-m^2)\,,
\eean
Then we obtain
\bean
\Xi(x)\eqn x(1+x)[(E^2-m^2)x+E^2-\Psi^2]\,.
\eean
We see immediately that as long as
\bean
E^2>m^2, \ \ \textrm{and}\ \  E^2>\Psi^2 \,,
\eean
$\Xi$ is positive everywhere. Therefore, two particles coming from infinity and colliding at $x=0$ could create infinite CM energies in this spacetime if $\Psi\neq 0$ for one of the particles. No fine-tuning is needed for either of the particles.

\section{Conclusion}
A comprehensive analysis on the BSW mechanism in five-dimensional Myers-Perry spacetims  was presented in this paper. We have discovered some important features which differ from those in four-dimensional Kerr spacetimes. By using both the radial and angular constraints, we show, for the first time, that nonextremal MP black holes cannot accelerate particles to arbitrarily high CM energies. For extremal black holes, we derive a general critical condition which is independent of the black hole parameters. When a naked singularity appears, we show that arbitrarily high CM energy can be created if one particle bounces back just at the singularity and collides with another ingoing particle. This requires fine-tuning on one particle's parameters, unlike the case in a four-dimensional Kerr black hole. A
special and interesting case is when the naked singularity just begins to form, because divergent CM energy can be produced even without fine-tuning. In this case, it is crucial that $\Psi$ for at least one particle is nonzero. Our results suggest that the BSW mechanism can help understand the natures of higher dimensional black holes and spacetimes with naked singularities.

\section*{Acknowledgements}
 This research was supported by NSFC Grants No. 11375026 and 11235003.

\end{document}